\documentstyle[aps,prl,twocolumn,epsf]{revtex}

\def\be{\begin{equation}}
\def\ee{\end{equation}}
\def\bea{\begin{eqnarray}}
\def\eea{\end{eqnarray}}
\def\sp{{\mathcal C}^2}

\newcommand{\la}{\langle}
\newcommand{\ra}{\rangle}

\begin{document}

\draft

\title{Three-qubit pure-state canonical forms}

\author{A. Ac\'{\i}n$^\dagger$, A. Andrianov$^{\dagger\ast}$, E. Jan\'e$^{\dagger}$ and\,R. Tarrach$^{\dagger}$}

\address{$^\dagger$Departament d'Estructura i Constituents de la
Mat\`eria, Universitat de Barcelona, Diagonal 647, E-08028 Barcelona, Spain.\\
$^\ast$Department of Theoretical Physics, St. Petersburg State University, 198904, St. Petersburg, Russia. \\
e-mail: acin@ecm.ub.es
}

\date{\today}

\maketitle


\begin{abstract}

In this paper we analyze the canonical forms into which any pure three-qubit state can be cast. 
The minimal forms, i.e. the ones with the minimal number of product states built from local bases, are also presented and lead to a complete classification of pure three-qubit states. This classification is related to the values of the polynomial invariants under local unitary transformations by a one-to-one correspondence.

\end{abstract}

\pacs{PACS Nos. 03.67.-a, 03.65.Bz}



\section{Introduction}

Non-local quantum correlations or entanglement between space-separated parties is one of the most fertile and thought-generating properties of quantum mechanics. Recently it has become a very useful resource for many of the applications in quantum information theory and this has led to a lot of work devoted to understanding how it can be quantified and manipulated.  

Bipartite pure state entanglement is almost completely understood, while many
questions are still open for the mixed state case. For pure states, the
Schmidt decomposition \cite{schmidt} has proven to be a very useful tool,
since it allows to write any pure state shared by two parties A and B in a
canonical form, where all the information about the non-local properties of
the state is contained in the positive Schmidt coefficients. The non-local
properties of quantum states can be also specified by means of other quantities invariant under the action of local unitary transformations. An interesting type of these invariants are given by polynomial combination of the coordinates of the state in a product basis, and the relation between these invariants and the Schmidt coefficients is well known.

Some novel aspects, compared to the bipartite case, appear for entangled
systems of more than two parties. In this work we study the canonical forms of three-qubit pure states, extending the results of bipartite systems. First we analyze the forms proposed for generalizing the Schmidt decomposition for three-qubit pure states. Then, we relate one of these decompositions to the polynomial invariants studied in \cite{pop,rains,GRB,LPS,kempe,sud,makhlin,nos}. We give a one-to-one correspondence between a canonical form for a three-qubit pure state and a complete set of polynomial invariants describing its entanglement properties. We also classify the different types of canonical forms by means of the minimal number of local bases product states (LBPS), i.e. the minimal number of non-local parameters, needed for the specification of a state. For any three-qubit pure state we give its decomposition with the minimal number of LBPS and the procedure that has to be applied in order to build it. Finally we indicate how to generalize the results to systems of $N$-qubits, where many difficulties arise.


\section{Generalization of the Schmidt decomposition}

The Schmidt decomposition has been a very useful tool for the study of entanglement properties of bipartite systems. For a generic bipartite pure state $|\Phi\ra\in{\mathcal C}^{d_1}\otimes{\mathcal C}^{d_2}$ it reads
\be
\label{bischmidt}
|\Phi\ra=\sum_{i=1}^l \alpha_i|ii\ra, \quad \alpha_i\geq 0,
\ee
where $l={\rm min}(d_1,d_2)$, $|ii\ra\equiv|i\ra_A \otimes |i\ra_B$, being $|i\ra$ orthonormal vectors in each subsystem, and $\alpha_i$ are the Schmidt coefficients.
It would be very interesting to find for three-qubit pure states a canonical
decomposition generalizing the features of the Schmidt decomposition. However,
the trivial generalization is not possible \cite{per} and it is not evident
how to extend the Schmidt decomposition to the case of $N$-party systems ($N>2$). Indeed several forms have been proposed (see for instance \cite{bc}).

In recent work \cite{nos} we gave a generalization of the Schmidt
decomposition for three-qubit pure states, in the sense that the coefficients
of this decomposition carry all the information about the non-local properties
of the state, and do so minimally and unambiguously, i.e. the decomposition is
not superflous.
Starting from a generic state shared by three parties, A, B and C,
\be
\label{state}
|\Psi\ra=\sum_{i,j,k}t_{ijk}|ijk\ra ,
\ee 
where $|ijk\ra\equiv |i\ra_A\otimes |j\ra_B\otimes |k\ra_C$, we look for the local bases that allow to write (\ref{state}) with the minimal number of LBPS. A simple counting of parameters shows that at least five product states built from local bases are needed in order to specify a generic state belonging to $\sp\otimes\sp\otimes\sp$. There are three inequivalent classes of five LBPS: the first one is the symmetric set
\be
\label{set1}
\{|000\ra,|001\ra,|010\ra,|100\ra,|111\ra\} ,
\ee
the second is weakly asymmetric and corresponds to the three sets of states,
\bea
\label{set2}
&&\{|000\ra,|001\ra,|100\ra,|110\ra,|111\ra\} \nonumber \\
&&\{|000\ra,|001\ra,|011\ra,|100\ra,|111\ra\} \nonumber \\
&&\{|000\ra,|010\ra,|100\ra,|101\ra,|111\ra\}  ,
\eea
and the third one is strongly asymmetric, and corresponds to the three sets
\bea
\label{set3}
&&\{|000\ra,|100\ra,|101\ra,|110\ra,|111\ra\} \nonumber \\
&&\{|000\ra,|010\ra,|011\ra,|110\ra,|111\ra\} \nonumber \\
&&\{|000\ra,|001\ra,|110\ra,|101\ra,|111\ra\} ,
\eea
where the three sets of states of (\ref{set2}) are related by permutation of the parties, and the same happens for the sets (\ref{set3}).
The non-equivalence between the sets (\ref{set1}), (\ref{set2}) and (\ref{set3}) follows from the different degrees of orthogonality between the five states within each set (see figure 1). In \cite{nos} it was proved that any three-qubit state can be written in terms of the product states of any of the asymmetric sets. Let us sketch the procedure.

Starting from a generic state (\ref{state}), one introduces the matrices $T_0
$ and $T_1 $ with elements
\be
\left(T_i\right)_{jk}\equiv t_{ijk}.
\ee
A change of basis on the first qubit transforms these matrices in the following way,
\bea
\label{u1}
T_0'&=&u_{00}^A T_0+u_{01}^A T_1 \nonumber \\
T_1'&=&u_{10}^A T_0+u_{11}^A T_1 ,
\eea
where $u_{ij}^A$ are the elements of a unitary matrix, while the effect of a change of basis in B (C) implies that each $T_i$ is left (right) multiplied by a unitary matrix $U^B(U^C)$.
The unitary transformation on party A is chosen such that
\be
\label{det}
{\rm det}(T_0')=0 .
\ee
There are always two solutions for this equation since (\ref{det}) is equivalent to
\be
\label{equis}
{\rm det}(T_0+xT_1)=0 ,
\ee
where $x\equiv\frac{u_{01}^A}{u_{00}^A}$ is an unbounded complex number. 
Now we apply two unitary matrices on parties B and C in order to diagonalize $T_0'$. These operations lead to the matrices
\bea
\label{matr3}
 M_0&\equiv& U^BT_0'U^C=\left(\matrix{\lambda_0 && 0 \cr 0 && 0}\right) \nonumber\\  
 M_1&\equiv& U^BT_1'U^C=\left(\matrix{\lambda_1 e^{i \varphi} && \lambda_2 \cr
     \lambda_3 && \lambda_4}\right ),
\eea
where $\lambda_i$ are real and positive, since all the phases have been absorbed by phase redefinitions of $|0\ra_A$, $|1\ra_A$, $|1\ra_B$ and $|1\ra_C$ . By means of these unitary transformations we have been able to write the initial state (\ref{state}) in terms of the products states appearing in the first set of (\ref{set3}), i.e.
\be
\label{schmidt}
|\Psi\ra=\lambda_0|000\ra+\lambda_1 e^{i \varphi}|100\ra+\lambda_2|101\ra+\lambda_3|110\ra+\lambda_4|111\ra .
\ee
Equation (\ref{equis}) has generically two different solutions, $x$ and $\bar
x$, so two different decompositions (\ref{schmidt}) are possible for the same
state $|\Psi\ra$. By limiting the range of the phase factor to
$0\leq\varphi\leq\pi$ a unique solution is found when $0<\varphi<\pi$ (see
\cite{nos} for more details), and then we have a unique canonical form in
which to cast almost any three-qubit pure state. For the remaining ones, when
$\varphi=0,\pi$, two canonical forms exist in general; we will break this
remaining degeneracy taking, for instance, the form with the smallest $\lambda
_1$, or, if $\lambda_1$ is unique, taking the form with the smallest $\lambda_0 $. It is important also to note that we have singled out party A in obtaining (\ref{schmidt}), but we could have chosen any of the three parties.

From (\ref{schmidt}) and by applying a unitary transformation on the third qubit,
\be
\label{u2}
|0'\ra=\frac{1}{\sqrt{\lambda_1^2+\lambda_2^2}}\left(\lambda_1 e^{i \varphi}|0\ra+\lambda_2|1\ra \right)
\ee
it follows that any state can be written, after removing the phases of four of the coefficients, as,
\be
|\Psi\ra=\eta_0 e^{i \phi}|000\ra+\eta_1|001\ra+\eta_2|100\ra+\eta_3|110\ra+\eta_4|111\ra ,
\ee
with $\eta_i$ real and positive, which corresponds to the first set in (\ref{set2}).


Recently, it has been shown \cite{hs} that the symmetric decomposition using
the set of states (\ref{set1}) is also possible. The proof is based on the
fact that if a given state $|\Psi\ra$ is written in a basis such that the state $|111\ra$ is the one that maximizes the overlap of $|\Psi\ra$ with any product state, i.e.
\be
\label{fid}
|t_{111}|^2={\rm max}\, |\la\Psi|\alpha\beta\gamma\ra|^2,
\ee
the coefficients $t_{110}$, $t_{101}$ and $t_{011}$ must be zero (otherwise one could find a product state with a larger overlap). Therefore  any state can be written as 


\be
\label{sud}
|\Psi\ra=\kappa_0e^{i \theta}|000\ra+\kappa_1|001\ra+\kappa_2|010\ra+\kappa_3|100\ra+\kappa_4|111\ra,
\ee
with $\kappa_i$ real and positive and $0\leq\theta<\pi$. Nevertheless, the conditions under which the decomposition (\ref{sud}) is unique are not known.


A different decomposition, which can also be thought as an alternative generalization of the Schmidt decomposition for three-qubit states, could be writing the state as a superposition of two product states, not necessarily orthogonal,
\be
\label{ghzdesc}
|\Psi\ra=\alpha|000\ra+\beta e^{i\delta}|\varphi_1\varphi_2\varphi_3\ra,   
\ee
with $\alpha$ and $\beta$ positive real numbers. This decomposition is only possible when $J_4\neq 0$ (see below for the definition of $J_4$) \cite{nos,ins}, which corresponds to the GHZ-class in \cite{ins}, and it has been proved very useful for the obtention of the optimal GHZ distillation protocol \cite{ghzdist}.


\section{The set of polynomial invariants}

The space of states of three qubits is $\sp\otimes\sp\otimes\sp$, which depends on sixteen real parameters (including the norm and the global phase). Two states, $|\Psi_1\ra$ and $|\Psi_2\ra$, are equivalent, as far as their entanglement properties are concerned, when they can be transformed one into the other by local unitary transformations.
Therefore the action of the elements of the group $U(1)\times SU(2)\times SU(2)\times SU(2)$ define orbits in the space of states, each orbit being the equivalence class of all the states having the same non-local properties. Thus, and as it is well-known, the dimension of a generic orbit for the case of three-qubit pure states is ten \cite{pop}, so six entanglement parameters should be enough to discriminate between two different orbits. Since the decomposition (\ref{schmidt}) is unique, it gives six quantities invariant under local unitaries, the five coefficients $\lambda_i$ and the phase $\varphi$, which allow us to check whether two generic states belong to the same orbit, i.e. whether they can be connected applying local unitary transformations. These parameters can be thought of as the entanglement coordinates. An alternative, though two-fold degenerate, set of entanglement parameters is given by polynomial combinations of the coefficients $t_{ijk}$ which are invariant under the group of local unitaries \cite{pop,rains,GRB,LPS,kempe,sud,makhlin}. In this section decomposition (\ref{schmidt}) will be related to these polynomial invariants.

For bipartite pure states, $|\Phi\ra\in{\mathcal C}^{d_1}\otimes{\mathcal
  C}^{d_2}$, a complete set of polynomial invariants, which allows to know whether two bipartite states have the same entanglement properties, is given by 
\be
{\rm tr}(\rho_A^l)={\rm tr}(\rho_B^l)\quad l=1,...,{\rm min}(d_1,d_2) ,
\ee
where $\rho_A\equiv{\rm tr}_B|\Phi\ra\la\Phi|$ and $\rho_B\equiv{\rm
  tr}_A|\Phi\ra\la\Phi|$ are the local density matrices. Since the eigenvalues of these matrices correspond to the square of the Schmidt coefficients (\ref{bischmidt}), we know the relation between the polynomial invariants and the Schmidt decomposition \cite{sud}. 

As it has been mentioned above, the space of entanglement parameters of pure
three-qubit states has dimension equal to six, so at least six linearly
independent polynomial combinations of $t_{ijk}$ invariant under local unitary transformations are needed in order to specify the non-local properties of a state, or the orbit which it belongs to. In \cite{sud} the six linearly independent polynomial invariants of minimal degree were found. The norm is a trivial one, so we will not consider it and in the rest of the paper we will restrict ourselves to the space of normalized states, the number of non-local parameters being reduced to five. This implies that we have $\sum_i\lambda_i^2=1$ in (\ref{schmidt}). Apart from the norm, the polynomial invariants given in \cite{sud} are
\bea
\label{is}
&&\frac{1}{2}\leq I_1\equiv {\rm tr}(\rho_A^2)\leq 1 \nonumber \\
&&\frac{1}{2}\leq I_2\equiv {\rm tr}(\rho_B^2)\leq 1 \nonumber \\
&&\frac{1}{2}\leq I_3\equiv {\rm tr}(\rho_C^2)\leq 1 \nonumber \\
&&\frac{1}{4}\leq I_4\equiv {\rm tr}(\rho_A\otimes\rho_B\rho_{AB})\leq 1 \nonumber \\
&&0\leq I_5\equiv|{\rm Hdet}(t_{ijk})|^2\leq\frac{1}{16},
\eea
where 
\bea
&&\rho_{A}\equiv {\rm tr}_{BC}|\Psi\rangle\langle\Psi| \nonumber\\
&&\rho_{B}\equiv {\rm tr}_{AC}|\Psi\rangle\langle\Psi| \nonumber\\
&&\rho_{C}\equiv {\rm tr}_{AB}|\Psi\rangle\langle\Psi| \nonumber\\
&&\rho_{AB}\equiv {\rm tr}_C|\Psi\rangle\langle\Psi| ,
\eea
and ${\rm Hdet}(t_{ijk})$ is the hyperdeterminant of the coefficients $t_{ijk}$ \cite{gel} and corresponds to the three-tangle of \cite{cof}. An equivalent set of invariants can be constructed \cite{nos}
\bea
\label{js}
&&J_1\equiv\frac{1}{4}(1+I_1-I_2-I_3-2\sqrt{I_5}) \nonumber \\
&&J_2\equiv\frac{1}{4}(1-I_1+I_2-I_3-2\sqrt{I_5}) \nonumber \\
&&J_3\equiv\frac{1}{4}(1-I_1-I_2+I_3-2\sqrt{I_5}) \nonumber \\
&&J_4\equiv \sqrt{I_5} \nonumber \\
&&J_5\equiv\frac{1}{4}(3-3I_1-3I_2-I_3+4I_4-2\sqrt{I_5}) , 
\eea
which, in terms of the parameters of the decomposition (\ref{schmidt}), are equal to
\bea
\label{inv}
&&0\leq J_1=|\lambda_1\lambda_4 e^{i \varphi}-\lambda_2\lambda_3|^2\leq\frac{1}{4} \nonumber \\
&&0\leq J_2=\mu_0\mu_2\leq\frac{1}{4} \nonumber \\
&&0\leq J_3=\mu_0\mu_3\leq\frac{1}{4} \nonumber \\
&&0\leq J_4=\mu_0\mu_4\leq\frac{1}{4} \nonumber \\
&&-\frac{1}{108}\leq J_5=\mu_0(J_1+\mu_2\mu_3-\mu_1\mu_4)\leq\frac{2}{27} ,
\eea
where $\mu_i \equiv \lambda_i^2$. It can be proved that $J_4$ and $J_5$ are invariant under permutation of the parties, because so are $2I_4-I_1-I_2$ and $I_5$, and $J_1$, $J_2$, and $J_3$ single out parties A, B and C respectively, and transform among themselves under party permutation.

From the above expressions one can prove the tighter bounds
\bea
\label{inv2}
&&0\leq J_2+J_3+J_4\leq\frac{1}{4}  \nonumber \\
&&0\leq J_1+J_3+J_4\leq\frac{1}{4} \nonumber \\
&&0\leq J_1+J_2+J_4\leq\frac{1}{4} \nonumber \\
&&0\leq J_4+J_5\leq\frac{1}{4} .
\eea
Also the following holds
\bea
\label{inv3}
&&J_1=0 \Rightarrow J_5=0  \nonumber \\
&&J_2=0 \Rightarrow J_5=0 \nonumber \\
&&J_3=0 \Rightarrow J_5=0 \nonumber \\
&&J_4=0 \Rightarrow \sqrt{J_1J_2J_3}=\frac{J_5}{2}. 
\eea

From (\ref{inv}) and using the normalization condition $\sum_i\mu_i=1$, it is possible to obtain the value of the set of coefficients $\{\mu_i\}$,
\bea
\label{mus}
&&\mu_0^\pm=\frac{J_4+J_5\pm\sqrt{\Delta_J}}{2(J_1+J_4)} \nonumber \\
&&\mu_i^\pm=\frac{J_i}{\mu_0^\pm},\quad i=2,3,4 \nonumber \\
&&\mu_1^\pm=1-\mu_0^\pm-\frac{J_2+J_3+J_4}{\mu_0^\pm},
\eea
where
\be
\label{discj}
\Delta_J\equiv(J_4+J_5)^2-4(J_1+J_4)(J_2+J_4)(J_3+J_4)\geq 0 ,
\ee
which implies
\be
\label{impl}
J_4+J_5=0 \Rightarrow J_4=J_5=0.
\ee
Note that the value of $\cos\varphi$ can be also found from (\ref{inv}),
\be
\label{cos}
\cos\varphi^\pm=\frac{\mu_1^\pm\mu_4^\pm+\mu_2^\pm\mu_3^\pm-J_1}{2\lambda_1^\pm\lambda_2^\pm\lambda_3^\pm\lambda_4^\pm} ,
\ee
and thus almost all the information about the decomposition can be extracted
from the values of the $\{J_i\}$. There remains however some ambiguity in
these expressions, since there are two solutions for the coefficients,
corresponding to $\mu_0^+$ and $\mu_0^-$, and for each of them, two different
angles, $0\leq\varphi^\pm\leq\pi$ and $\tilde \varphi^\pm=2\pi-\varphi^\pm $,
coming from (\ref{cos}). Part of this uncertainty is due to the two solutions
of (\ref{det}) and in fact the coefficients $\{\mu_i^+,\varphi^+\}$ and
$\{\mu_i^-,\tilde\varphi^-\}$ describe the same orbit, and the same happens
for $\{\mu_i^-,\varphi^-\}$ and $\{\mu_i^+,\tilde\varphi^+\}$. As it has been
said, the solutions associated to $\tilde\varphi$ are not considered because
of the range of the angle. However the set of invariants $\{J_i\}$ (or
$\{I_i\}$) does not determine a unique orbit, or equivalently a canonical
point representing it. Two candidates are possible,
$\{\mu_i^\pm,\varphi^\pm\}$, so there is still some ambiguity left.

The five polynomial invariants (\ref{is}) are real, and this means that they
can not distinguish among the orbits associated to a given pure three-qubit
state $|\Psi\ra$, with coefficients $t_{ijk}$, and to $|\Psi\ra^\ast$, given by $t_{ijk}^\ast$. Indeed,
\be
\label{amb}
I_i(|\Psi\ra^\ast)=I_i(|\Psi\ra)^\ast=I_i(|\Psi\ra) ,
\ee
where the second equality comes from the fact that the invariants are real. It
is not possible, due to this ambiguity, to individuate a unique canonical
state representing an orbit from the invariants (\ref{is}), or (\ref{js}). A
twelfth degree complex polynomial invariant, $I_6$, introduced by Grassl \cite{grassl}, solves (albeit redundantly) this problem, just by inspection of the sign of its imaginary part (in other words, the second equality of (\ref{amb}) is not valid for this invariant). The explicit form of Grassl's invariant, using decomposition (\ref{schmidt}) is
\be
\label{grassl}
I_6=\mu_0^2\mu_4(\lambda_4(1-2(\mu_0+\mu_1))+2\lambda_1\lambda_2\lambda_3 e^{-i\varphi})^2 .
\ee
The set given by (\ref{is}) and $I_6$ is complete, it allows to check when two states belong to different orbits, and from their values one can obtain a unique canonical point representing the orbit applying (\ref{mus}-\ref{cos}) and, in the end, using $I_6$ to discriminate between the two candidates.

This situation is quite different from what happens for pure states of
bipartite systems. In this case, a generic state $|\Phi\ra\in{\mathcal
  C}^{d_1}\otimes{\mathcal C}^{d_2}$, with coefficients $t_{ij}$, can be always tranformed into $|\Phi\ra^\ast$ by local unitary transformations, as this is clear from the fact that all the Schmidt coefficients are real. In general this is not true for three-qubit systems, although in some cases the state $|\Psi\ra$ and its complex conjugate $|\Psi\ra^\ast$ are in the same orbit. This corresponds to the situations when either
\be
|\cos\varphi^+|=|\cos\varphi^-|=1 ,
\ee
or
\bea
\cos\varphi^+&=&\cos\varphi^- \nonumber\\
\mu_i^+&=&\mu_i^-.
\eea
Equivalent conditions in terms of the invariants $\{J_i\}$ can be obtained, giving
\be
\label{real}
\sqrt{J_1J_2J_3}=\frac{|J_5|}{2} ,
\ee
for the first case and
\be
\label{mueq}
\Delta_J=0 ,
\ee
for the second. Furthermore in both situations a product basis can be found
for which all the coefficients $t_{ijk}$ are real. For the states satisfying the first condition, this basis is the one that gives decomposition (\ref{schmidt}), since we have $e^{i\varphi}=\pm1$, while in the second case the proof is a bit more tedious and it is given in the appendix A. From these results, then, it follows that
\be
\label{complex}
|\Psi\ra\sim|\Psi\ra^\ast\Leftrightarrow\sqrt{J_1J_2J_3}=\frac{|J_5|}{2}\,\,{\rm or}\,\,\Delta_J=0\Leftrightarrow|\Psi\ra\,{\rm real} ,
\ee
where a pure state belonging to $\sp\otimes\sp\otimes\sp$ is said to be real when there exists a product basis where all the coefficients are real.

To summarize, five independent quantities invariant under local unitaries are
needed to specify the non-local properties of a generic three-qubit pure
state. The coefficients appearing in the decomposition (\ref{schmidt}) form a
complete faithful and minimal set of such invariants, when constrained as
explained after (\ref{schmidt}). The polynomial invariants given in (\ref{is})
must be completed with $I_6$ in order to solve the ambiguity between the orbits associated to $|\Psi\ra$ and $|\Psi\ra^\ast$, and from the values of these polynomial invariants one can build a unique canonical point representing the orbit. Also when $|\Psi\ra$ and $|\Psi\ra^\ast$ are in the same orbit there exists a product basis where all the coordinates of $|\Psi\ra$ are real, as it happens for pure states of bipartite systems. 

Let us mention finally that any real state can be written with real
coefficients in terms of a set of six LBPS, adding the state $|011\rangle$ to (\ref{set1}) or to the first of (\ref{set3}). This is done by diagonalizing $T_0$ with two orthogonal matrices.

\section{Minimal decomposition}

We have seen that a generic three-qubit pure state can always be written in
terms of five product states from any of the sets of states in (\ref{set1}),
(\ref{set2}) or (\ref{set3}). However it is not clear which set should be used
to find the minimal decomposition, that is, the one with the least number of
non-local parameters. The minimal number of LBPS needed to specify a state
$|\Psi\ra$ will be denoted by $\nu(\Psi)$. We know that in general $\nu=5$ but
now we want to analyze the cases in which $\nu<5$. In this section we present
a complete classification of the three-qubit pure states according to this
minimal number of product states. We also give necessary and sufficient
conditions written in terms of the invariants $\{J_i\}$ to be satisfied by the
states of each class. The number of non-local parameters in each family is
$\nu-1$, since all the coefficients are real. All the families satisfy
condition (\ref{real}).

\subsection {$\nu=4$}

There are several subfamilies of states that allow for a decomposition in terms of four LBPS.

\medskip

{\bf Type 4a}: This subfamily is given by the states with $\mu_4=0$ in (\ref{schmidt}). It is easy to prove that this condition is equivalent to $J_4=0$ (we will take the rest of invariants different from zero, unless otherwise specified). Condition (\ref{real}) is also satisfied with $J_5>0$, since all the phases can be absorbed.

{\bf Type 4b}: States with $\mu_2=0$ ($\mu_3=0$) in (\ref{schmidt}). The equivalent conditions in term of the invariants are $J_2=J_5=0$ ($J_3=J_5=0$). Let us mention that there is an apparently lack of symmetry in this subfamily, but this is due to the fact that party A has been singled out in the determinations of the decomposition (\ref{schmidt}). In fact the analogous states with $J_1=J_5=0$ are written with four terms if either party B or C is singled out in (\ref{u1}-\ref{matr3}).

{\bf Type 4c}: States with $\mu_1=0$ in (\ref{schmidt}). It can be proved that the corresponding conditions in terms of the invariants are $J_1J_4+J_1J_2+J_1J_3+J_2J_3=\sqrt{J_1J_2J_3}=\frac{J_5}{2}$. Again the lack of symmetry is due to the fact that party A is privileged in the calculation of the decomposition (\ref{schmidt}). Analogous condition can be found interchanging the role of the indices 1, 2 and 3, which means that the minimal decompositions is obtained if one of the other two parties is singled out in (\ref{u1}-\ref{matr3}).

{\bf Type 4d}: States with $\kappa_0=0$ in (\ref{sud}). It is proved in appendix B that the corresponding condition, apart from (\ref{real}), which is always satisfied when $\nu<5$, is $\Delta_J=0$.

\subsection{$\nu=3$}

Now we move to the study of those states that can be expressed as a sum of three LBPS.

\medskip

{\bf Type 3a}: This subfamily is given by taking $\mu_1=\mu_4=0$ in (\ref{schmidt}). The equivalent conditions for the invariants are $J_4=0$ and $J_1J_2+J_1J_3+J_2J_3=\sqrt{J_1J_2J_3}=\frac{J_5}{2}$.

{\bf Type 3b}: These states correspond to the case $\mu_j=\mu_k=0$ in
(\ref{schmidt}), for $j,k\in \{1,2,3\}$ and $j\neq k$. These conditions expressed in terms of the invariants are $J_j=J_k=J_5=0$.

\subsection{$\nu=2$}

The states with two product states built from local bases are just in two classes.

\medskip

{\bf Type 2a}: $J_i=0$ except $J_1 (J_2,J_3)$, and these are the states where party A(B,C) is not entangled with the other two parties, so there is not truly three-party entanglement.

{\bf Type 2b}: $J_i=0$ except $J_4$, they include the standard GHZ state.

\subsection{$\nu=1$}

\medskip

{\bf Type1}: $J_i=0$, and these are the product states where there is no correlation between the parties.

\subsection{Summary}

All the states belonging to $\sp\otimes\sp\otimes\sp$ have been classified in terms of the minimal number, $\nu$, of LBPS required to express the state, and the resulting families of states are shown in table \ref{class}. Generically five terms are needed, although there are cases where $\nu<5$. Necessary and sufficient conditions in terms of the set of invariants $\{J_i\}$ are given, which can be used to recognise the subfamily a three-qubit pure state belongs to. Once this has been done, we have provided the procedure that has to be applied in order to find this minimal decomposition with product states. 

\section{Generalization to more parties}

The decomposition (\ref{schmidt}), which generalizes the bipartite Schmidt decomposition, has been proved to be very fruitful for the case of three-qubit pure states, so it will be convenient to know the way it can be generalized to more parties. In this section first we will consider with some details the case of four-qubit systems and this will give us insight into the difficulties found when we try to extend our results.

The procedure to be applied for the generalization of decomposition
(\ref{schmidt}) for pure states belonging to
$\sp\otimes\sp\otimes\sp\otimes\sp$, i.e. states
$|\Psi\ra=\sum_{i,j,k,l}t_{ijkl}|ijkl\ra$ shared by four parties A, B, C and D, will be now described. First we define the two hypermatrices \cite{gel}
\be
(T_i)_{jkl}\equiv t_{ijkl} ,
\ee
which means that the initial state is interpreted as
\be
|\Psi\ra=|0\ra|\phi_0\ra+|1\ra|\phi_1\ra ,
\ee
where $|\phi_i\ra$ are, up to normalization, three-qubit pure states, their coordinates being given by the elements of the corresponding hypermatrix $T_i$. The effect of the change of local bases is very similar to the one described for three-qubit systems: a unitary transformation on system A mixes the coordinates of the two $|\phi_i\ra$, while unitary transformations on the rest of subsystems can be used to make zero some of their coefficients.
Now we apply the change of local bases on A that gives
\be
\label{hdet}
{\rm Hdet}(T_0')=0 ,
\ee
and afterwards unitary transformation on B, C and D are used to write the new
$|\phi_0'\ra$ in the canonical decomposition found for three-qubit pure states. Since (\ref{hdet}) is verified, it is known that $|\phi_0'\ra$ belongs to, at least, type 4a states, so we will manage to write the initial state $|\Psi\ra$ in terms of the twelve product states:
\bea
\label{4qset}
&&|0000\ra,|0100\ra,|0101\ra,|0110\ra, \nonumber\\
&&|1000\ra,|1001\ra,|1010\ra,|1011\ra, \nonumber\\
&&|1100\ra,|1101\ra,|1110\ra,|1111\ra .
\eea

A simple counting of parameters gives that the minimal number of LBPS needed
to specify a state $|\Psi\ra\in\sp\otimes\sp\otimes\sp\otimes\sp$ is exactly twelve. The decomposition we have found depends on twenty-four non-local parameters but it is known that by phase redefinitions, i.e. acting locally with $U(1)$, five phases can be absorbed (generically, for $N$ parties $N+1$ coefficients can be made real), so the number of non-local parameters is nineteen (including the norm), as it was expected \cite{pop}.

However some problems arise in this case. Many decompositions in terms of the
set of states (\ref{4qset}) are possible for the same state. In fact
(\ref{hdet}) is a fourth degree equation, so four solutions will be found and
from these solutions four different decompositions will be derived. For the
case of three-qubit pure state there were two solutions for (\ref{det}), but
we managed to obtain a unique decomposition by limiting the range of
$\varphi$. A similar reasoning seems not to be trivial for this
case. Furthermore for pure four-qubit states more inequivalent set of twelve
product states appear, and this will difficult the analysis of the minimal
decomposition. The generalization of decomposition (\ref{schmidt}) to
$N$-qubit pure states ($N>3$) is then quite cumbersome.

Finally, it has to be noted that the algorithm proposed in \cite{hs} for the decomposition (\ref{sud}) can be also extended to higher dimensional systems. Let us mention however that, in any case, as the dimension of the space increases, the number of coefficients that can be made equal to zero in any of the decompositions becomes irrelevant.

\section{Conclusions}

In this work we have studied the canonical forms of pure three-qubit states, extending the known results of bipartite systems. 

First we show the possible generalizations of the Schmidt decomposition and we
relate one of these decompositions (\ref{schmidt}) to the polynomial
invariants of \cite{pop,rains,GRB,LPS,kempe,sud,makhlin,nos}. The six linearly
independent polynomial invariants of \cite{sud} are not able to discriminate
betwee the entanglement orbits associated to a state and its complex conjugate
in a product basis. An additional polynomial invariant introduced in
\cite{grassl} has to be used, and we have seen how to connect this complete
set of polynomial invariants with our generalization of the Schmidt
decomposition. Indeed it is shown how to find a canonical point in a generic
orbit described by this complete set of invariants. Let us mention here that a
three-qubit pure state $|\Psi\ra$ and its complex conjugate $|\Psi\ra^\ast$ give the same optimal probability of distilling a maximally entangled state of three qubits, in the single-copy case \cite{ghzdist}.

We have also looked for the decomposition of any state, $|\Psi\ra$, with the minimal number, $\nu(\Psi)$, of product states built from local bases. Generically this number is equal to five, although many exceptional states have been found with $\nu<5$. We have been able to give a complete classification of these states by means of a set of necessary and sufficient conditions written in terms of the set of invariants (\ref{js}). The procedure to be applied in order to build the minimal decomposition for every state has been given too. The classification of the pure three-qubit states in terms of their entanglement properties can be done following alternative criteria to the one described here, which is based on the features observed acting with the group of local unitary transformations. A possible approach is to classify the states looking for their probabilistic conversions under local operations and classical comunication (LOCC) for the single-copy case (see \cite{ins} and also \cite{ghzdist,bc2}) or in the asymptotic regime \cite{asym}. It would be expected that these classifications are a coarse-graining of the one presented in this work. In fact this is the case for the equivalences classes under LOCC given in \cite{ins}.

Finally it has been indicated how to extend decomposition (\ref{schmidt}) to
systems of more parties. A simple counting of parameters shows that at least
$2^N-N$ product states built from local bases are needed in order to specify a generic $N$-qubit pure state, and for four qubits we succeeded to find a procedure that makes zero four of the coordinates $t_{ijkl}$. The decomposition (\ref{sud}) allows for a simpler generalization. However, in all the cases some difficulties arise, related to the uniqueness of the decompositions, and it is not clear whether these generalized Schmidt decompositions are quite useful for composite systems of more than three qubits.

\section*{Appendix A: Real states}

In this appendix we will show that, given a pure three-qubit state
$|\psi\ra\in\sp\otimes\sp\otimes\sp$, this state is real, i.e. there exists a product basis for which all coefficients are real, if and only if $\sqrt{J_1J_2J_3}=\frac{|J_5|}{2}$ or $\Delta_J=0$.

Consider the case of a state $|\psi\ra=\sum_{i,j,k}t_{ijk}|ijk\ra$ where all
the $t_{ijk}$ are real. Now we will follow the procedure described by the equations (\ref{u1}-\ref{matr3}) that gives us the decomposition (\ref{schmidt}). Since the initial coordinates are real, from (\ref{det}) a second degree equation in $x$ with real coefficients is obtained, and this implies that the two solutions, $x$ and $\bar x$, satisfy that either they are both real or $x=\bar x^\ast$. In the first case, the calculation of the decomposition can be performed using orthogonal matrices, and since the initial coordinates were real, we will obtain a real decomposition, i.e. $\varphi=0,\pi$, which is equivalent to (\ref{real}). For the second case, since $x=\bar x^\ast$, ${\rm tr}(T_0'^{\dagger}T_0')={\rm tr}(\bar T_0'^{\dagger}\bar T_0')$, and then $\mu_0=\bar \mu_0$ and (\ref{mueq}) is satisfied.

Now, the inverse has to be proved. For the first case it is clear that all the states verifying (\ref{real}) take real coordinates when they are expressed in the basis used in decomposition (\ref{schmidt}). For the second case the proof is not so trivial.

Consider a generic state, $|\phi\ra$, having $\Delta_J$ equal to zero. The parametrization of this family of states is simplified using (\ref{ghzdesc}), so let us first mention some facts about this decomposition. As it has been shown, any state with $J_4\neq 0$ can be written as (\ref{ghzdesc}) \cite{nos,ins} where
\bea
\label{ghzpar}
\alpha&=&\frac{1}{\lambda_4}\sqrt{J_1+J_4} \nonumber\\
\beta&=&\frac{1}{\lambda_4}\sqrt{\mu_2\mu_3+\mu_4(\mu_4+\mu_2+\mu_3)} \nonumber\\
\delta&=&\arg(\lambda_1\lambda_4e^{i\varphi}-\lambda_2\lambda_3) ,
\eea
and, up to unitary transformations,
\be
|0\ra=\left(\matrix{1 \cr 0}\right)\qquad|\varphi_i\ra=\left(\matrix{\cos\gamma_i \cr \sin\gamma_i}\right),\, i=1,2,3 .
\ee
It can be proved that when $\Delta_J=0$ the coefficients $\alpha$ and $\beta$ are equal and then the states to be studied are
\be
\label{djst}
|\phi\ra=\alpha\left(|000\ra+e^{i\delta}|\varphi_1\varphi_2\varphi_3\ra\right) .
\ee
Recall that for these states the complex conjugate is in the same orbit as the original one, and this means that
\be
\label{compl}
t_{ijk}^\ast=\sum v_{ia}^1v_{jb}^2v_{kc}^3t_{abc} ,
\ee
where $t_{ijk}$ are the coordinates in some product basis and $v_{ia}^1$, $v_{jb}^2$ and $v_{kc}^3$ are the elements of the local unitary matrices, $V^1$ in A, $V^2$ in B and $V^3$ in C, connecting the two states. From (\ref{djst}) it follows that these unitary operators are
\be
\label{unop}
 V^i=e^{-i\delta'}\left(\matrix{c_i && s_i \cr s_i && -c_i}\right) ,
\ee
where $c_i\equiv\cos\gamma_i$, $s_i\equiv\sin\gamma_i$ and $\delta'\equiv\frac{\delta}{3}$ (actually, the phase factors in the matrices $V^i$ can be given by arbitrary angles $\delta_i$ satisfying the constraint $\sum_i\delta_i=\delta$, but we choose these angles for simplicity).

Now we would like to find a product basis for which all the coefficients are real, i.e.
\be
t_{ijk}'=\sum w_{ia}^1w_{jb}^2w_{kc}^3t_{abc}=t_{ijk}'^\ast ,
\ee
and from this condition and using (\ref{compl}), we have
\be
V^i=(W^i)^TW^i .
\ee
The explicit form of each $V^i$, (\ref{unop}), as a product of a phase factor and a real and symmetric matrix allows to write them as
\be
\label{undiag}
V^i=e^{-i\delta'}(O^i)^TD^iO^i ,
\ee
where $O^i$ are orthogonal matrices and $D^i$ are diagonal matrices with entries $\pm 1$. The change of basis we are looking for then is given by
\be
\label{ws}
W^i=(D^i)^{\frac{1}{2}}O^i=e^{-i\delta''}\left(\matrix{\tilde c_i && \tilde s_i \cr -i\tilde s_i && i\tilde c_i}\right) ,
\ee
where $\tilde c_i\equiv\cos\frac{\gamma_i}{2}$, $\tilde
s_i\equiv\sin\frac{\gamma_i}{2}$ and $\delta''\equiv\frac{\delta}{6}$. The new coordinates obtained applying these local change of basis are, up to normalization,
\bea
\label{treal}
&&t_{000}'= \phantom{-}\tilde c_1\tilde c_2\tilde c_3\cos\frac{\delta}{2}\quad t_{001}'= -\tilde c_1\tilde c_2\tilde s_3\sin\frac{\delta}{2} \nonumber\\
&&t_{010}'= -\tilde c_1\tilde s_2\tilde c_3\sin\frac{\delta}{2}\quad t_{011}'= -\tilde c_1\tilde s_2\tilde s_3\cos\frac{\delta}{2} \nonumber\\
&&t_{100}'= -\tilde s_1\tilde c_2\tilde c_3\sin\frac{\delta}{2}\quad t_{101}'= -\tilde s_1\tilde c_2\tilde s_3\cos\frac{\delta}{2} \nonumber\\
&&t_{110}'= -\tilde s_1\tilde s_2\tilde c_3\cos\frac{\delta}{2}\quad t_{111}'= \phantom{-}\tilde s_1\tilde s_2\tilde s_3\sin\frac{\delta}{2} .
\eea
This ends the proof.

\section*{Appendix B: Type 4d}

In this section we prove that a three-qubit pure state $|\psi\ra$ can be written as
\be
\label{4d}
|\psi\ra=l_1|001\ra+l_2|010\ra+l_3|100\ra+l_4|111\ra ,
\ee
with real and positive coefficients, if and only if (\ref{real}) and (\ref{mueq}) are verified.

Starting from (\ref{4d}) we can apply the procedure given by
(\ref{u1}-\ref{matr3}) to obtain (\ref{schmidt}). It can be seen that all the
unitary matrices needed for the determination of this decomposition are real,
i.e. they are orthogonal, and since the original coefficients $\{l_i\}$ were also real, we will obtain a real decomposition with (\ref{real}). Moreover, it can also be proved that the two matrices obtained after (\ref{det}), $T_0'$ and $\bar T_0'$, corresponding to the two solutions of this equation, $x$ and $\bar x$, verify 
\be
{\rm tr}((T_0')^{\dagger}T_0')={\rm tr}((\bar T_0')^{\dagger}\bar T_0') .
\ee
This condition implies that $\mu_0=\bar \mu_0$, and using (\ref{mus}) we have also (\ref{mueq}).

Now we prove the inverse. Consider a state $|\phi\ra$ satisfying (\ref{real}) and (\ref{mueq}). Because of the latter condition, the state allows for a decomposition as (\ref{djst}). Moreover, since (\ref{real}) is also satisfied, we have $\varphi=0,\pi$ in (\ref{schmidt}), and this implies, using (\ref{ghzpar}), that $\delta=0,\pi$. The generic expression for a state satisfying both the conditions can be now given,
\be
\label{4dgen}
|\phi\ra=\alpha(|000\ra\pm|\varphi_1\varphi_2\varphi_3\ra).
\ee
If we perform the local change of bases described by (\ref{ws}) it can be
seen, using (\ref{treal}) and the fact that $\delta=0,\pi$, that the state $|\phi\ra$ is of type 4d. Indeed, the new coordinates are, after absorbing the phases and up to normalization,
\be
l_1=\tilde s_1\tilde s_2\tilde c_3\quad l_2=\tilde s_1\tilde c_2\tilde s_3\quad l_3=\tilde c_1\tilde s_2\tilde s_3\quad l_4=\tilde c_1\tilde c_2\tilde c_3 ,
\ee
for $\delta=0$, and
\be
l_1=\tilde c_1\tilde c_2\tilde s_3\quad l_2=\tilde c_1\tilde s_2\tilde c_3\quad l_3=\tilde s_1\tilde c_2\tilde c_3\quad l_4=\tilde s_1\tilde s_2\tilde s_3 ,
\ee
for $\delta=\pi$. Note that the local bases that appear in (\ref{4d}) are the ones that diagonalize the local density matrices. This gives the procedure to be applied in order to find the minimal decomposition without performing the maximization of (\ref{fid}), which is generically a more difficult calculation.

\bigskip

\section*{Acknowledgments}
We thank G. Vidal, A. Sudbery and M. Grassl for useful discussion.
R. T. acknowledges financial support by CICYT project AEN 98-0431, CIRIT project 1998SGR-00026 and CEC project IST-1999-11053, A. Andrianov by RFBR 99-01-00736 and CIRIT, PIV-2000, A. Ac\'{\i}n and E. J. by a grant from MEC (AP98 and AP99).
Financial support from the ESF is also acknowledged. This work was partially performed at the 2000 Benasque Center for Science.


\newpage

\begin{figure}
\begin{center}
 \epsffile{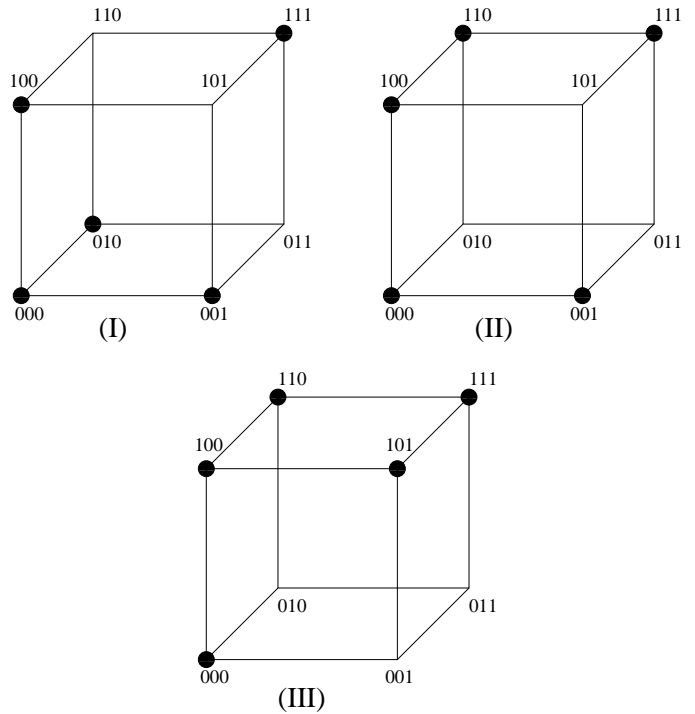}
\medskip
\caption{The figure depicts the three inequivalent sets of states given by (\ref{set1}), (\ref{set2}) and (\ref{set3}).}
\end{center}
\end{figure}

\begin{table}
\begin{tabular}{|c|c|c|}
  Type & Conditions & States \\ 
\hline
  4a & $J_4=0,\sqrt{J_1J_2J_3}=\frac{J_5}{2}$ & $|000\ra,|100\ra,|101\ra,|110\ra$ \\
\hline
  4b & $J_2=J_5=0$ & $|000\ra,|100\ra,|110\ra,|111\ra$ \\
\hline
  4c & $J_1J_4+J_1J_2+J_1J_3+J_2J_3=$ & $|000\ra,|101\ra,|110\ra,|111\ra$ \\
  & $\sqrt{J_1J_2J_3}=\frac{J_5}{2}$ & \\
\hline
  4d & $\Delta_J=0,\sqrt{J_1J_2J_3}=\frac{|J_5|}{2}$ & $|001\ra,|010\ra,|100\ra,|111\ra$ \\
\hline
  3a & $J_1J_2+J_1J_3+J_2J_3=$ & $|000\ra,|101\ra,|110\ra$ \\
  & $\sqrt{J_1J_2J_3}=\frac{J_5}{2},J_4=0$ & \\
\hline
  3b & $J_1=J_2=J_5=0$ & $|000\ra,|110\ra,|111\ra$ \\
\hline
  2a & All $J_i=0$ apart from $J_1$ & $|000\ra,|011\ra$ \\
\hline
  2b & All $J_i=0$ apart from $J_4$ & $|000\ra,|111\ra$ \\
\hline 
  1 & $J_i=0$ & $|000\ra$ \\
\end{tabular}
\medskip
\caption{Classification of three-quantum-bit states. For the types of states denoted by 4b, 4c, 3b and 2a, there exist analogous condition interchanging the roles of the invariants $J_1,J_2,J_3$, and consequently the product states used in the minimal decomposition.}
\label{class}
\end{table}

\end{document}